\documentclass[12pt]{article}
\usepackage{amsmath}
\usepackage{graphicx}
\usepackage{amsfonts}
\usepackage{amssymb}

\newcommand{\beq}{\begin{equation}}
\newcommand{\eeq}{\end{equation}}
\newcommand{\beqa}{\begin{eqnarray}}
\newcommand{\eeqa}{\end{eqnarray}}

\begin{document}
%
\begin{titlepage}
\rightline{DSF-32/2000}
\rightline{NORDITA-2000/94 HE}
\vskip1.2cm
\centerline{\Large\bf Off-shell tachyon amplitudes:}
\centerline{\Large\bf analyticity and projective invariance\footnote
{Research partially supported by the EC  RTN programme  HPRN-CT-2000-00131 in which R.M. is associated with Nordita and the other authors with  Frascati-LNF.}%
}
\vskip1.2cm
\centerline{\bf
Francesco Cuomo$^a$, Raffaele Marotta$^b$,  Francesco Nicodemi$^a$,}
\centerline{\bf
Roberto Pettorino$^a$, Franco Pezzella$^a$ and Gianluca Sabella$^a$}
\centerline{\sl
$ ^a$ Dipartimento di Scienze Fisiche, Universit\`a di Napoli "Federico II"}
\centerline{\sl and I.N.F.N., sezione di Napoli}
\centerline{\sl
Complesso Universitario di Monte S. Angelo, Via Cintia, I-80126 Napoli, Italy}
\centerline{\sl name.surname@na.infn.it}
\centerline{\sl$ ^b$ NORDITA, Blegdamsvej 17, DK-2100 Copenhagen \O, Denmark }
\centerline{\sl name.surname@nordita.dk}
\vskip2cm
\begin{abstract}
We compute off-shell three- and four-tachyon amplitudes at tree level by using a prescription based on the
requirement of projective invariance.  In particular we show that the off-shell four tachyon amplitude can be put in the same form as the corresponding
on-shell one,
exhibiting therefore the same analyticity properties. This is shown both for the bosonic and for
the fermionic string.  The result obtained in the latter case can be extended to the off-shell four-tachyon amplitude in type 0 theory.
\end{abstract}
\end{titlepage}

\newpage

\section{Introduction}

The presence of the tachyon in the spectrum of string theories has been always
considered as a disease even though it was generally seen as a sign of a wrong
choice of the vacuum. Superstring theory, while on the one hand has solved the
problems of the physical spectrum projecting out tachyons, on the other hand
has obscured the interest in the dynamics of these particles. In the last
years the proposal by A. Sen \cite{S} of a well defined mechanism for tachyon
condensation has attracted wide interest \cite{tachcond}. The main ingredient
of this scenario is the tachyon potential which depends on the formulation of
the off-shell theory. Several different approaches have been followed in order
to provide such an extension (see e.g. \cite{CMNP}), many of which in the
context of string field theory \cite{W}. We think that some help in
understanding these features may be provided by the operator formalism of the
$N$-string Vertex \cite{dV} for evaluating tachyon amplitudes suitably
extended off-shell.

In some recent papers \cite{CMPP} \cite{LMP} prescriptions have been given in
order to compute off-shell string scattering amplitudes in the above mentioned
formalism. In particular in \cite{LMP} it has been proposed a prescription
based on the property of projective invariance that must be exhibited both by
on- and off-shell amplitudes. Their being projective invariant is crucial if
factorization is required to hold \cite{KV}. Such a requirement inspires the
right choice for the local coordinate systems defined around the punctures of
the external states. In fact the $N$-string Vertex is an operator which
depends on $N$ complex Koba-Nielsen variables, corresponding to those
punctures, through $N$ projective transformations $V_{i}(z)$ which define
local coordinate systems vanishing around each $z_{i}$, i.e.
\[
V_{i}(0)=z_{i}.
\]

The $N$-string Vertex can be regarded as a sort of functional generator of
scattering amplitudes among arbitrary string states. When it is saturated with
$N$ physical on-shell states, the corresponding amplitude is independent of
the $V_{i}$'s while, when saturated on off-shell states, its dependence on\/
these maps is transferred to the amplitudes themselves. Hence, on-shell string
amplitudes are independent of the choice of such local coordinate systems
around the punctures and are described in terms of correlation functions of
vertex operators which are primary fields of the underlying conformal field
theory. On the contrary, off-shell string amplitudes depend on the choice of
the local coordinate systems and are given by correlation functions of vertex
operators corresponding to quasi-primary fields. There is an analogy with
gauge theories. Indeed choosing $V_{i}(z)$ is equivalent to perform a gauge
choice, since on-shell amplitudes are gauge invariant, while their off-shell
counterparts are not.

In the case of off-shell string amplitudes not all the choices of the $V_{i}%
$'s are equivalent; requiring their projective invariance suggests the
suitable ones.

In this paper, following the above prescription, we compute off-shell
amplitudes involving three and four tachyons starting from the bosonic string
case. In particular we show that in the case of open strings, the off-shell
amplitude can be put in the same form as its on-shell counterpart, so
recovering the same analytic properties. This result will be shown to be valid
also for the closed string if one enforces the kinematic condition
\begin{equation}
s+t+u=4m^{2} \label{kincond}%
\end{equation}
where $s=-(p_{1}+p_{2})^{2},t=-(p_{1}+p_{4})^{2},u=-(p_{1}+p_{3})^{2}$ are the
usual Mandelstamm variables and $m$ is the tachyon mass, but $p_{i}^{2}$ is unconstrained.

The computation has been then extended to the fermionic string case and of
course limited to the four-tachyon amplitude, since the three-tachyon one is
vanishing. In this case one has to take into consideration a dependence of the
off-shell amplitudes on the \emph{pictures} which must be assigned to the
tachyons in order to saturate the ghost number conservation on the sphere.
Projective invariance again inspires the choice of the local coordinate
systems around the punctures. We show that while for a given picture
assignment the off-shell amplitudes depend on the performed choice, the
average over all possible picture assignments yields an off-shell four-tachyon
amplitude of the same form as the on-shell one. Also this result follows after
imposing the kinematic constraint (\ref{kincond}).

The paper is organized as follows.

In sect 2. we report the results relative to the off-shell three and
four-tachyon amplitudes in the bosonic open and closed string theory.

In sect. 3 we apply the same procedure to the four-tachyon amplitude in the
fermionic string case. We notice that such a computation is the same as in
type 0 theory.

\setcounter{equation}{0}

\section{Three- and four-tachyon bosonic string amplitudes}

\subsection{Open bosonic string}

We start from the $N$-string $0$-loop vertex $\widehat{V}_{N;0}^{op}$
generating the tree diagrams of oriented open strings \cite{DPFHLS} and
specialized to the case of $N$ tachyons. It assumes the following form:%

\begin{equation}
\widehat{V}_{N;0}^{op}=C_{0}^{op}\left\langle \Omega\right|  \int\left[
dm\right]  _{N}^{0}\prod_{i=1}^{N}\exp\left\{  \alpha^{\prime}\ln
V_{i}^{\prime}\left(  0\right)  \hat{p}_{i}^{2}\right\}  \prod
_{\substack{i,j=1 \\i<j }}^{N}\exp\left\{  2\alpha^{\prime}\lg(z_{i}%
-z_{j})\hat{p}_{i}\cdot\hat{p}_{j}\right\}  ,
\end{equation}
where $C_{0}^{op}$ is a normalization factor given, in $d$ dimensions, by
\cite{DMLRM}:%

\begin{equation}
C_{0}^{op}=g_{o}^{-2}\frac{1}{\left(  2\alpha^{^{\prime}}\right)  ^{d/2}},
\label{c0open}%
\end{equation}
$g_{o}$ being the open string coupling constant. The bra $\left\langle
\Omega\right|  \equiv\Pi_{i=1}^{N}$ $\left\langle x_{i}=0\right|  $
$\delta\left(  \sum_{i=1}^{N}\hat{p}_{i}\right)  $ represents the product of
the vacua of the Fock spaces of each tachyon. The measure is defined by%

\begin{equation}
\lbrack dm]_{N}^{0}=\frac{1}{dV_{abc}}\prod_{i=1}^{N-1}\left[  \vartheta
\left(  z_{i}-z_{i+1}\right)  \right]  \prod_{i=1}^{N}\frac{dz_{i}}%
{V_{i}^{\prime}\left(  0\right)  } \label{open measure}%
\end{equation}
and $dV_{abc}$ is the projective invariant volume element%

\[
dV_{abc}=\frac{dz_{a}dz_{b}dz_{c}}{\left(  z_{a}-z_{b}\right)  \left(
z_{b}-z_{c}\right)  \left(  z_{a}-z_{c}\right)  }.
\]
Let us denote by $\left|  p\right\rangle $ a tachyon state with momentum $p$.
It is created by the vertex operator%

\[
\mathcal{V}(z)=\mathcal{N}_{t}:e^{i\sqrt{2\alpha^{^{\prime}}}p\cdot X(z)}:
\]
where the colons denote the standard normal ordering on the modes of the open
string coordinate $X^{\mu}(z)$ and $\mathcal{N}_{t}$ is a normalization factor
\cite{DMLRM}:%

\[
\mathcal{N}_{t}=2g_{o}\left(  2\alpha^{\prime}\right)  ^{(d-2)/4}.
\]
If we write, as usual,%

\[
X^{\mu}(z)=\hat{x}^{\mu}-i\hat{p}^{\mu}\log z+i\sum_{n\neq0}\frac{\hat{\alpha
}^{\mu}}{n}z^{-n}
\]
then the tachyon state is%

\[
\left|  p\right\rangle \equiv\lim_{z\rightarrow0}\mathcal{V}_{p}(z)\left|
0;p=0\right\rangle =\mathcal{N}_{t}e^{ip\cdot\hat{x}}\left|
0;p=0\right\rangle .
\]
The tachyon is on-shell if
\[
p^{2}=-m^{2}=\frac{1}{\alpha^{\prime}}.
\]
Saturating $\widehat{V}_{N;0}^{op}$ with $N$ off-shell tachyons, with momenta
$p^{2}\neq\frac{1}{\alpha^{\prime}}$, one gets the following contribution to
the off-shell $N$-tachyon amplitude:

\smallskip%

\begin{align}
\mathcal{A}_{N}^{op}(p_{1},\cdot\cdot\cdot,p_{N})  &  =C_{0}^{op}%
\mathcal{N}_{t}^{N}\int\frac{1}{dV_{abc}}\prod_{i=1}^{N}\left[  dz_{i}%
(V_{i}^{\prime}(0))^{\alpha^{\prime}p_{i}^{2}-1}\right] \nonumber\\
&  \times\prod_{i=1}^{N-1}\vartheta(z_{i}-z_{i+1})\prod_{\substack{i,j=1 \\i<j
}}^{N}(z_{i}-z_{j})^{2\alpha^{\prime}p_{i}\cdot p_{j}}.
\label{offshellNtachyon}%
\end{align}

The total amplitude is to be obtained by summing over non-cyclic permutations
of the external states.

\subsubsection{Projective invariance}

While for on-shell amplitudes the dependence on the local coordinate systems
$V_{i}$ clearly disappears, this isn't true for the off-shell ones, which
depend on them. Requiring projective invariance for the off-shell amplitudes
inspires the choice of the $V_{i}$'s. Under an $SL(2,\mathcal{R})$
transformation
\[
z\rightarrow z^{\prime}=\frac{az+b}{cz+d}\,\,\,\,\,\,\,\,ad-bc\neq0\quad
a,b,c,d\in\mathcal{R},
\]
the quantities $V_{i}^{\prime}(0)$ have to transform according to
\[
V_{i}^{\prime}(0)\rightarrow V_{i}^{\prime}(0)\frac{ad-bc}{(cz_{i}+d)^{2}}.
\]
A suitable choice is then
\begin{equation}
V_{i}^{\prime}(0)=\frac{(z_{i-1}-z_{i})(z_{i}-z_{i+1})}{(z_{i+1}-z_{i-1})}%
\rho(z_{1},...,z_{N}), \label{vi(0)}%
\end{equation}
with $z_{0}$ $=$ $z_{N}$, $z_{N+1}$ $=$ $z_{1}$ and $\rho$ any projective
invariant function of the punctures.

In the following we will choose $\rho=1,$ in which case the expression in
(\ref{vi(0)}) coincides with the first derivative evaluated in $z=0$ of the
so-called Lovelace function used in the dual models \cite{L}. Under an
$SL(2,\mathcal{R})$ transformation, which preserves the cycling ordering, the
other factors in the amplitude (\ref{offshellNtachyon}) transform as
\begin{equation}
\left\{
\begin{array}
[c]{c}%
dz_{i}\rightarrow\frac{ad-bc}{\left(  cz_{i}+d\right)  ^{2}}dz_{i}\\
z_{i}-z_{j}\rightarrow\frac{ad-bc}{(cz_{i}+d)(cz_{j}+d)}(z_{i}-z_{j})\\
dV_{abc}\rightarrow dV_{abc}%
\end{array}
\right.  . \label{projective variations}%
\end{equation}
Therefore the integrand $\mathcal{I}$ in the amplitude transforms according
to
\[
\mathcal{I}\rightarrow\prod_{i=1}^{N}\left\{  \left[  \frac{ad-bc}{\left(
cz_{i}+d\right)  ^{2}}\right]  ^{\alpha^{\prime}p_{i}^{2}-1}\times\left[
\frac{ad-bc}{\left(  cz_{i}+d\right)  ^{2}}\right]  ^{-\alpha^{\prime}%
p_{i}^{2}}\right\}  \mathcal{I},
\]
while the measure transforms as
\[
\prod_{i=1}^{N}dz_{i}\rightarrow\prod_{i=1}^{N}\left[  \frac{ad-bc}{\left(
cz_{i}+d\right)  ^{2}}\right]  dz_{i},
\]
showing that the off-shell amplitude is projective invariant.

In computing amplitudes we are therefore allowed to fix three of the punctures
at three specific points on the real axis.

\subsubsection{Three- and four-tachyon amplitudes}

For the three-tachyon case the integral in (\ref{offshellNtachyon}) is just
one and the amplitude turns out to be the product of $C_{0}^{op}$ by
$\mathcal{N}_{t}^{3}$, independently of the tachyon momenta\footnote{With a
different choice in (\ref{vi(0)}) of $\rho$ (which in the case $N=3$ is just a
constant), a dependence on the momenta would appear in the amplitude via a
factor $\rho^{\sum_{i}\alpha^{\prime}p_{i}^{2}-3}$.}, i.e.:
\begin{equation}
\mathcal{A}_{3}=8g_{o}(2\alpha^{\prime})^{(d-6)/4}. \label{3tachyonamp}%
\end{equation}

\smallskip

In the case of the four-tachyon amplitude, if we fix the M\"{o}bius gauge as
$z_{1,2,4}=\infty,1,0$ the contribution to the amplitude results to be

\smallskip%

\begin{align}
\mathcal{A}_{4}^{op}  &  =C_{0}^{op}\mathcal{N}_{t}^{4}\int_{0}^{1}%
dzz^{-\alpha^{\prime}s-2}(1-z)^{-\alpha^{\prime}t-2}\nonumber\\
&  =16g_{0}^{2}(2\alpha^{\prime})^{(d-4)/2}B(-\alpha^{\prime}s-1,-\alpha
^{\prime}t-1). \label{4op}%
\end{align}

The total off-shell four-tachyon amplitude is obtained by considering the sum
over the independent non cyclic permutations:
\begin{align}
\mathcal{A}_{4}^{op}(s,t,u)  &  =16g_{o}^{2}(2\alpha^{\prime})^{(d-4)/2}%
\left\{  \frac{\Gamma[-\alpha(s)]\Gamma[-\alpha(t)]}{\Gamma[-\alpha
(s)-\alpha(t)]}+\frac{\Gamma[-\alpha(s)]\Gamma[-\alpha(u)]}{\Gamma
[-\alpha(s)-\alpha(u)]}\right. \nonumber\\
&  \quad\quad\quad\quad\quad\quad\quad\left.  +\frac{\Gamma[-\alpha
(t)]\Gamma[-\alpha(u)]}{\Gamma[-\alpha(t)-\alpha(u)]}\right\}  , \label{a4fin}%
\end{align}
being $\alpha(x)=\alpha^{\prime}x+1$ the usual Regge trajectory. This
expression is identical to the Veneziano amplitude, but here the tachyon
momenta are only restricted to verify momentum conservation.

It is trivial to check that the residue of (\ref{a4fin}) at the tachyon pole
is just $\left(  A_{3}\right)  ^{2},$ as it should.

\subsection{Closed string}

The $N$-tachyon vertex for the closed string reads \cite{DPFHLS}
\[
\widehat{V}_{N;0}^{cl}=C_{0}^{cl}\left\langle \Omega\right|  \int\left[
dm\right]  _{N}^{0}\exp\left\{  \frac{1}{2}\sum_{i=1}^{N}\frac{\alpha^{\prime
}}{2}\hat{p}_{i}^{2}\ln\left|  V_{i}^{\prime}\left(  0\right)  \right|
^{2}\right\}
\]
\begin{equation}
\times\exp\left\{  \sum_{\substack{i,j=1 \\i\neq j }}^{N}\frac{\alpha^{\prime
}}{2}\hat{p}_{i}\cdot\hat{p}_{j}\ln\left|  z_{i}-z_{j}\right|  \right\}  ,
\label{vclosed}%
\end{equation}
where the measure is%

\[
\left[  dm\right]  _{N}^{0}=\frac{1}{dV_{abc}}\prod_{i=1}^{N}\frac{d^{2}z_{i}%
}{\left|  V_{i}^{\prime}(0)\right|  ^{2}}
\]
and the $SL(2,C)$ invariant volume element is:%

\[
dV_{abc}=\frac{d^{2}z_{a}d^{2}z_{b}d^{2}z_{c}\ }{\left|  z_{a}-z_{b}\right|
^{2}\left|  z_{a}-z_{c}\right|  ^{2}\left|  z_{c}-z_{b}\right|  ^{2}}.
\]
$C_{0}^{cl}$ is the overall normalization for tree closed string amplitudes.
It has the same expression as in the open string case (\ref{c0open}), $g_{o}$
being now substituted by the closed string coupling constant that we will
indicate by $g_{c}$. Analogously to the open string case, we define a tachyon
state with momentum $p$ as follows:%

\[
\left|  p\right\rangle =\lim_{z,\bar{z}\rightarrow0}\mathcal{V}_{p}(z,\bar
{z})\left|  0,p=0\right\rangle =\mathcal{N}_{t}e^{ip\cdot\widehat{x}}\left|
0,p=0\right\rangle
\]
with $\mathcal{N}_{t}$ being a normalization factor given by \cite{DMLRM}:%

\[
\mathcal{N}_{t}=2\sqrt{2\pi}g_{c}\left(  2\alpha^{\prime}\right)  ^{(d-2)/4}.
\]
Saturating (\ref{vclosed}) with $N$ tachyon states gives the following
result:
\begin{equation}
\mathcal{A}_{N}^{cl}(p_{1},...,p_{N})=C_{0}^{cl}\mathcal{N}_{t}^{N}\int
\frac{1}{dV_{abc}}\prod_{i=1}^{N}\left[  d^{2}z_{i}\left|  V_{i}^{\prime
}(0)\right|  ^{\frac{\alpha^{\prime}}{2}p_{i}^{2}-2}\right]  \prod
_{\substack{i,j=1 \\i<j }}^{N}\left|  z_{i}-z_{j}\right|  ^{\alpha^{\prime
}p_{i}\cdot p_{j}}. \label{offshellNtachyoncl}%
\end{equation}
Once again for on-shell tachyons $\left(  p_{i}^{2}=\frac{4}{\alpha^{^{\prime
}}}\right)  $ the local maps cancel out. For arbitrary momenta the same choice
of the $V_{i}^{\prime}(0)$'s made in the open string case guarantees that the
integrand in (\ref{offshellNtachyoncl}) is still $SL(2,C) $ invariant.

For $N=3$ the expression in (\ref{offshellNtachyoncl}) specifies in%

\begin{equation}
\mathcal{A}_{3}^{cl}=8(2\pi)^{3/2}g_{c}(2\alpha^{\prime})^{(d-6)/4}.
\label{3tachyonclosed}%
\end{equation}
Again this is independent of the tachyon momenta.

With the usual choice of the punctures the four tachyon amplitude reads
\begin{equation}
\mathcal{A}_{4}^{cl}=C_{o}^{cl}\mathcal{N}_{t}^{4}\int d^{2}z\left|  z\right|
^{-\frac{\alpha^{\prime}}{2}s-4}\left|  1-z\right|  ^{-\frac{\alpha^{\prime}%
}{2}t-4}. \label{4tachyonclosed}%
\end{equation}
The leading Regge trajectory for closed strings is
\[
\alpha\left(  s\right)  =\frac{\alpha^{\prime}}{2}s+2
\]
so, in terms of the Euler beta function, we can write
\begin{equation}
\mathcal{A}_{4}^{cl}=(2\pi)^{2}16g_{c}^{2}(2\alpha^{\prime})^{(d-4)/2}B\left(
-\frac{1}{2}\alpha\left(  s\right)  ,-\frac{1}{2}\alpha\left(  t\right)
,\frac{1}{2}\alpha\left(  s\right)  +\frac{1}{2}\alpha\left(  t\right)
+1\right)  . \label{4tachcl}%
\end{equation}
If we restrict to the kinematic shell
\begin{equation}
s+t+u\equiv-\sum_{i=1}^{4}p_{i}^{2}=-\frac{16}{\alpha^{\prime}},
\label{kinematic shell}%
\end{equation}
we find that
\[
\frac{1}{2}\alpha\left(  s\right)  +\frac{1}{2}\alpha\left(  t\right)
=-\frac{1}{2}\alpha\left(  u\right)  -1
\]
so
\begin{equation}
\mathcal{A}_{4}^{cl}=(2\pi)^{2}16g_{c}^{2}(2\alpha^{\prime})^{(d-4)/2}B\left(
-\frac{1}{2}\alpha\left(  s\right)  ,-\frac{1}{2}\alpha\left(  t\right)
,-\frac{1}{2}\alpha\left(  u\right)  \right)  . \label{4tachclfin}%
\end{equation}
This, again, is the same formula one obtains in the on-shell case, with the
difference that now the tachyon momenta are only constrained to verify
momentum conservation and (\ref{kinematic shell}).

\setcounter{equation}{0}

\section{Four-tachyon fermionic string amplitudes}

\subsection{Open fermionic string}

In the case of the fermionic string the vertex operator corresponding to a
physical state is not unique. If fact one can associate to each physical state
an infinite set of vertex operators corresponding to different values of the
ghost number, or, equivalently, corresponding to different picture numbers.
Nevertheless physical quantities like on-shell scattering amplitudes must be
independent of the picture assignment. In order to ensure ghost number
conservation on the sphere the picture numbers of the scattering strings must
add up to -2. In this paper we will only consider picture numbers -1 and 0.
The vertex operator for tachyons in picture 0 is
\[
\mathcal{V}_{p}^{(0)}(z)=p\cdot\psi(z)e^{ip\cdot X(z)},
\]
while in picture -1 one has
\[
\mathcal{V}_{p}^{(-1)}(z)=e^{-\phi(z)}e^{ip\cdot X(z)}.
\]
The scalar field $\phi(z)$ has a simple expansion in oscillators given by
\[
\phi(z)=\hat{x}+\hat{N}\lg z+\sum_{n\neq0}\frac{\hat{\alpha}_{n}}{n}z^{-n},
\]
where
\[
\left[  \hat{x},\hat{N}\right]  =1;\quad\quad\quad\left[  \hat{\alpha}%
_{m},\hat{\alpha}_{n}\right]  =-m\delta_{m+n,0}.
\]
The zero mode acts on a state in the picture $a$ according to
\[
\hat{N}\left|  \chi\right\rangle _{a}=-a\left|  \chi\right\rangle _{a}.
\]

The $N$-tachyon Vertex to be used in this case can be obtained from the
$N$-string Vertex for the Neveu-Schwarz string in Ref. \cite{CP} with a new
factor taking into account the contribution of the scalar $\phi$ field
\cite{DPFHLS}. The resulting vertex is
\begin{align}
\widehat{V}_{N;0}^{op}  &  =\left\langle \Omega\right|  \int\left[  dm\right]
_{N}^{0}\exp\left\{  -\alpha^{\prime}\sum_{\substack{i,j=1\\i\neq j}}^{N}%
\ln\frac{\sqrt{V_{i}^{\prime}\left(  0\right)  V_{j}^{\prime}\left(  0\right)
}}{z_{i}-z_{j}}\hat{p}_{i}\cdot\hat{p}_{j}\right\} \nonumber\\
&  \times\exp\left\{  -\frac{i}{2}\sum_{\substack{i,j=1\\i\neq j}}^{N}%
\frac{\sqrt{V_{i}^{\prime}\left(  0\right)  V_{j}^{\prime}\left(  0\right)  }%
}{z_{i}-z_{j}}b_{1/2}^{(i)}\cdot b_{1/2}^{(j)}-\frac{1}{2}\sum
_{\substack{i,j=1\\i\neq j}}^{N}\ln\frac{\sqrt{V_{i}^{\prime}\left(  0\right)
V_{j}^{\prime}\left(  0\right)  }}{z_{i}-z_{j}}\hat{N}_{i}\hat{N}_{j}\right\}
, \label{Vnclosos}%
\end{align}
where the measure is defined in (\ref{open measure}). Notice that here, as in
the definition of the tachyon state, we skip the normalization factors which
are different from the bosonic case.

It is straightforward to see that saturating $\widehat{V}_{N;0}^{op}$ with
$N=3$ tachyons yields zero for any value of the tachyon momenta.

On the other hand, the contribution to the amplitude obtained by saturating
this vertex on a four-tachyon state $\left|  \Omega^{\prime}\right\rangle $,
with the tachyons put in some picture P$_{i}$, depends on the particular
picture chosen. For example if we choose to put the four tachyons in the
picture P$_{1}$=[0,0,-1,-1] we have
\[
\left|  \Omega^{\prime}\right\rangle =p_{1}\cdot b_{-1/2}^{(1)}\left|
0;p_{1}\right\rangle _{0}\otimes p_{2}\cdot b_{-1/2}^{(2)}\left|
0;p_{2}\right\rangle _{0}\otimes\left|  0;p_{3}\right\rangle _{-1}%
\otimes\left|  0;p_{4}\right\rangle _{-1}
\]
and the corresponding amplitude is%

\begin{align}
\mathcal{A}_{4}^{op(1)}(p_{1},...,p_{4})  &  =p_{1}\cdot p_{2}\int\frac
{1}{dV_{abc}}\prod_{i=1}^{3}\left[  \vartheta\left(  z_{i}-z_{i+1}\right)
\right]  \prod_{i=1}^{4}\left[  dz_{i}V_{i}^{\prime}\left(  0\right)
^{\alpha^{\prime}p_{i}^{2}-\frac{1}{2}}\right] \nonumber\\
&  \times\prod_{i<j=1}^{4}(z_{i}-z_{j})^{2\alpha^{\prime}p_{i}\cdot p_{j}%
}\frac{1}{(z_{1}-z_{2})(z_{3}-z_{4})}. \label{ampfs}%
\end{align}
In the fermionic case the on-shell condition for open string tachyons reads
\[
p_{i}^{2}=\frac{1}{2\alpha^{\prime}}
\]
so the local map cancellation in the on-shell amplitude is again trivially verified.

It is easy to show that projective invariance for the off-shell amplitude
still holds here if $V_{i}^{\prime}(0)$ is given by (\ref{vi(0)})$.$ Indeed
the expression (\ref{ampfs}) only differs from the bosonic one by a factor
\[
\frac{\prod_{i=1}^{4}V_{i}^{\prime}(0)^{\frac{1}{2}}}{(z_{1}-z_{2}%
)(z_{3}-z_{4})},
\]
which can be shown to be projective invariant making use of the relations
(\ref{projective variations}). With the usual choice for the fixed punctures
we can write the amplitude (\ref{ampfs}) as
\begin{align}
\mathcal{A}_{4}^{op(1)}(p_{1},...,p_{4})  &  =p_{1}\cdot p_{2}\int_{0}%
^{1}dzz^{-\alpha^{\prime}s-2}\left(  1-z\right)  ^{-\alpha^{\prime}%
t-1}\label{a4op}\\
&  =p_{1}\cdot p_{2}\frac{\Gamma(-\alpha^{\prime}s-1)\Gamma(-\alpha^{\prime
}t)}{\Gamma(-\alpha^{\prime}s-\alpha^{\prime}t-1)},\nonumber
\end{align}
which manifestly depends on the picture assignment. Furthermore this amplitude
exhibits an unwanted singularity for $s=-1/\alpha^{\prime}$. On-shell one has
\[
2\alpha^{\prime}p_{1}\cdot p_{2}=-\alpha^{\prime}s-1
\]
and the kinematic factor just cancels the unwanted singularity, but off-shell
this is not the case. The solution to this puzzle is that the contribution to
the amplitude in each channel is to be obtained by averaging over all possible
picture assignments. Then one recovers the expected analyticity properties if
condition (\ref{kincond}) is imposed.

There are six different possible choices of the picture assignments and the
corresponding amplitudes are gathered into couples, each of which yielding the
same contribution. For example $\mathcal{A}_{4}^{op(1)}$ in (\ref{a4op}) is
coupled with the amplitude
\[
\mathcal{A}_{4}^{op(2)}=p_{3}\cdot p_{4}\frac{\Gamma(-\alpha^{\prime
}s-1)\Gamma(-\alpha^{\prime}t)}{\Gamma(\alpha^{\prime}u+1)}
\]
corresponding to P$_{2}$=[-1,-1,0,0], to give
\[
\mathcal{A}_{4}^{op(1)}+\mathcal{A}_{4}^{op(2)}=\left(  p_{1}\cdot p_{2}%
+p_{3}\cdot p_{4}\right)  \frac{\Gamma(-\alpha^{\prime}s-1)\Gamma
(-\alpha^{\prime}t)}{\Gamma(\alpha^{\prime}u+1)}.
\]
The kinematic condition (\ref{kincond}) now reads
\begin{equation}
s+t+u=-\frac{8}{\alpha^{\prime}} \label{kinematic shell2}%
\end{equation}
ensuring that
\[
p_{1}\cdot p_{2}+p_{3}\cdot p_{4}=-s-\frac{1}{\alpha^{\prime}},
\]
by which we obtain
\[
\mathcal{A}_{4}^{op(1)}+\mathcal{A}_{4}^{op(2)}\sim\frac{\Gamma(-\alpha
^{\prime}s)\Gamma(-\alpha^{\prime}t)}{\Gamma(\alpha^{\prime}u+1)}.
\]
Here we have used the standard formula
\[
a\Gamma(a)=\Gamma(a+1).
\]
The remaining four possible pictures give rise to identical contributions and
the total off-shell scattering amplitude can therefore be written as
\begin{equation}
\mathcal{A}_{4}^{op}(s,t,u)\sim\left[  \frac{\Gamma(-\alpha^{\prime}%
s)\Gamma(-\alpha^{\prime}t)}{\Gamma(\alpha^{\prime}u+1)}+\frac{\Gamma
(-\alpha^{\prime}s)\Gamma(-\alpha^{\prime}u)}{\Gamma(\alpha^{\prime}%
t+1)}+\frac{\Gamma(-\alpha^{\prime}t)\Gamma(-\alpha^{\prime}u)}{\Gamma
(\alpha^{\prime}s+1)}\right]  . \label{a4offshellf}%
\end{equation}
Also in the fermionic case, therefore, the off-shell four-tachyon amplitude
assumes the same form as the corresponding on-shell one. The formula
(\ref{a4offshellf}) is valid for arbitrary $s,t,u$ provided they satisfy
(\ref{kinematic shell2}).

\subsection{Closed fermionic string}

The $N$-tachyon vertex can be obtained duplicating the open string vertex:%

\begin{align}
\widehat{V}_{N;0}^{cl}  &  =\left\langle \Omega\right|  \int\left[  dm\right]
_{N}^{0}\exp\left\{  \frac{1}{2}\sum_{\substack{i,j=1 \\i\neq j }}^{N}%
\frac{\alpha^{\prime}}{2}\ln\frac{\left|  z_{i}-z_{j}\right|  ^{2}}{\left|
V_{i}^{\prime}\left(  0\right)  V_{j}^{\prime}\left(  0\right)  \right|  }%
\hat{p}_{i}\cdot\hat{p}_{j}\right\} \nonumber\\
&  \times\exp\left\{  -\frac{i}{2}\sum_{\substack{i,j=1 \\i\neq j }}^{N}%
\frac{\sqrt{V_{i}^{\prime}\left(  0\right)  V_{j}^{\prime}\left(  0\right)  }%
}{z_{i}-z_{j}}b_{1/2}^{(i)}\cdot b_{1/2}^{(j)}-\frac{i}{2}\sum
_{\substack{i,j=1 \\i\neq j }}^{N}\frac{\sqrt{\bar{V}_{i}^{\prime}\left(
0\right)  \bar{V}_{j}^{\prime}\left(  0\right)  }}{\bar{z}_{i}-\bar{z}_{j}%
}\tilde{b}_{1/2}^{(i)}\cdot\tilde{b}_{1/2}^{(j)}\right\} \nonumber\\
&  \times\exp\left\{  \frac{1}{2}\sum_{\substack{i,j=1 \\i\neq j }}^{N}%
\ln\frac{\sqrt{V_{i}^{\prime}\left(  0\right)  V_{j}^{\prime}\left(  0\right)
}}{z_{i}-z_{j}}N_{i}\cdot N_{j}+\frac{1}{2}\sum_{\substack{i,j=1 \\i\neq j
}}^{N}\ln\frac{\sqrt{\bar{V}_{i}^{\prime}\left(  0\right)  \bar{V}_{j}%
^{\prime}\left(  0\right)  }}{\bar{z}_{i}-\bar{z}_{j}}\tilde{N}_{i}\cdot
\tilde{N}_{j}\right\}  . \label{Vnclosfs}%
\end{align}
In this case there are of course 36 possible choices for the picture numbers,
which must add to -2 in each sector. Let us consider for example the choice
made in \cite{KT}, that is P$_{1}$=[(0,0),(0,0),(-1,-1),(-1,-1)]. The
corresponding amplitude is obtained by saturating (\ref{Vnclosfs}) on the
state
\[
\left|  \Omega^{\prime}\right\rangle =p_{1}\!\cdot b_{-1/2}^{(1)}p_{1}%
\!\cdot\tilde{b}_{-1/2}^{(1)}\left|  0;p_{1}\right\rangle _{0,0}\otimes
p_{2}\!\cdot b_{-1/2}^{(2)}p_{2}\!\cdot\tilde{b}_{-1/2}^{(2)}\left|
0;p_{2}\right\rangle _{0,0}\otimes\left|  0;p_{3}\right\rangle _{-1,-1}%
\otimes\left|  0;p_{4}\right\rangle _{-1,1}.
\]
This results in the following expression
\begin{align}
\mathcal{A}_{4}^{cl(1)}(p_{1},...,p_{4})  &  =(p_{1}\cdot p_{2})^{2}\int
\frac{1}{dV_{abc}}\prod_{i=1}^{4}\left[  d^{2}z_{i}\left|  V_{i}^{\prime
}\left(  0\right)  \right|  ^{\alpha^{\prime}\frac{p_{i}^{2}}{2}-1}\right]
\nonumber\\
&  \times\prod_{\substack{i,j=1 \\i<j }}^{4}\left|  z_{i}-z_{j}\right|
^{\alpha^{\prime}p_{i}\cdot p_{j}}\frac{1}{\left|  z_{1}-z_{2}\right|
^{2}\left|  z_{3}-z_{4}\right|  ^{2}}.
\end{align}
The independence of the local coordinate maps is still trivially verified in
the on-shell case, i.e. when
\[
p_{i}^{2}=\frac{2}{\alpha^{\prime}},
\]
while projective invariance holds also off-shell, as in the open string case,
provided that the $V_{i}^{\prime}(0)$'s are again given by (\ref{vi(0)}).

With the standard choice for the punctures the amplitude assumes the form
\begin{align*}
\mathcal{A}_{4}^{cl(1)}  &  =(p_{1}\cdot p_{2})^{2}\int d^{2}z\left|
z\right|  ^{-\frac{\alpha^{\prime}}{2}s-4}\left|  1-z\right|  ^{-\frac
{\alpha^{\prime}}{2}t-2}\\
&  =(p_{1}\cdot p_{2})^{2}B\left(  -\frac{\alpha^{\prime}}{4}s-1,-\frac
{\alpha^{\prime}}{4}t,\frac{\alpha^{\prime}}{4}(s+t)+2\right)  .
\end{align*}
Forcing the tachyon momenta to live on the kinematic shell
\begin{equation}
s+t+u\equiv-\sum_{i=1}^{4}p_{i}^{2}=-\frac{8}{\alpha^{\prime}},
\label{closed kinematic shell}%
\end{equation}
the amplitude can be rewritten as
\begin{align*}
\mathcal{A}_{4}^{cl(1)}  &  =(p_{1}\cdot p_{2})^{2}B\left(  -\frac
{\alpha^{\prime}}{4}s-1,-\frac{\alpha^{\prime}}{4}t,-\frac{\alpha^{\prime}}%
{4}u\right) \\
&  =\pi(p_{1}\cdot p_{2})^{2}\frac{\Gamma(-\frac{\alpha^{\prime}}{4}%
s-1)\Gamma(-\frac{\alpha^{\prime}}{4}t)\Gamma(-\frac{\alpha^{\prime}}{4}%
u)}{\Gamma(\frac{\alpha^{\prime}}{4}s+2)\Gamma(\frac{\alpha^{\prime}}%
{4}t+1)\Gamma(\frac{\alpha^{\prime}}{4}u+1)}.
\end{align*}
Just as in the open string case, the result strongly depends on the picture
assignment and has the wrong analyticity properties. However there exist three
more pictures yielding the same ratio of $\Gamma$ functions, namely P$_{2}%
$=[(-1,-1),(-1,-1),(0,0),(0,0)], P$_{3}$=[(0,-1),(0,-1), (-1,0),(-1,0)] and
P$_{4}$=[(-1,0),(-1,0),(0,-1),(0,-1)], the corresponding amplitudes being
\begin{align*}
\mathcal{A}_{4}^{cl(2)}  &  =\pi(p_{3}\cdot p_{4})^{2}\frac{\Gamma
(-\frac{\alpha^{\prime}}{4}s-1)\Gamma(-\frac{\alpha^{\prime}}{4}%
t)\Gamma(-\frac{\alpha^{\prime}}{4}u)}{\Gamma(\frac{\alpha^{\prime}}%
{4}s+2)\Gamma(\frac{\alpha^{\prime}}{4}t+1)\Gamma(\frac{\alpha^{\prime}}%
{4}u+1)},\\
\mathcal{A}_{4}^{cl(3)}  &  =\pi(p_{1}\cdot p_{2})(p_{3}\cdot p_{4}%
)\frac{\Gamma(-\frac{\alpha^{\prime}}{4}s-1)\Gamma(-\frac{\alpha^{\prime}}%
{4}t)\Gamma(-\frac{\alpha^{\prime}}{4}u)}{\Gamma(\frac{\alpha^{\prime}}%
{4}s+2)\Gamma(\frac{\alpha^{\prime}}{4}t+1)\Gamma(\frac{\alpha^{\prime}}%
{4}u+1)},\\
\mathcal{A}_{4}^{cl(4)}  &  =\pi(p_{1}\cdot p_{2})(p_{3}\cdot p_{4}%
)\frac{\Gamma(-\frac{\alpha^{\prime}}{4}s-1)\Gamma(-\frac{\alpha^{\prime}}%
{4}t)\Gamma(-\frac{\alpha^{\prime}}{4}u)}{\Gamma(\frac{\alpha^{\prime}}%
{4}s+2)\Gamma(\frac{\alpha^{\prime}}{4}t+1)\Gamma(\frac{\alpha^{\prime}}%
{4}u+1)}.
\end{align*}
Summing the four kinematic factors and enforcing (\ref{closed kinematic
shell}) leads to
\[
(p_{1}\cdot p_{2}+p_{3}\cdot p_{4})^{2}=(s+\frac{4}{\alpha^{\prime}})^{2}%
\]
so that
\begin{align*}
\mathcal{A}_{4}^{cl}  &  =\mathcal{A}_{4}^{cl(1)}+\mathcal{A}_{4}%
^{cl(2)}+\mathcal{A}_{4}^{cl(3)}+\mathcal{A}_{4}^{cl(4)}\\
&  \sim\frac{\Gamma(-\frac{\alpha^{\prime}}{4}s)\Gamma(-\frac{\alpha^{\prime}%
}{4}t)\Gamma(-\frac{\alpha^{\prime}}{4}u)}{\Gamma(\frac{\alpha^{\prime}}%
{4}s+1)\Gamma(\frac{\alpha^{\prime}}{4}t+1)\Gamma(\frac{\alpha^{\prime}}%
{4}u+1)}.
\end{align*}
Similar considerations can be used to deal with contributions arising from the
other 32 possible picture assignments. They all can be seen to belong to
groups of four contributions adding up to the same expression, so that the
total amplitude is found to be
\begin{equation}
\mathcal{A}_{4}^{cl}(s,t,u)\sim\frac{\Gamma(-\frac{\alpha^{\prime}}{4}%
s)\Gamma(-\frac{\alpha^{\prime}}{4}t)\Gamma(-\frac{\alpha^{\prime}}{4}%
u)}{\Gamma(\frac{\alpha^{\prime}}{4}s+1)\Gamma(\frac{\alpha^{\prime}}%
{4}t+1)\Gamma(\frac{\alpha^{\prime}}{4}u+1)}. \label{a4cl}%
\end{equation}
The amplitude (\ref{a4cl}) coincides with the corresponding on-shell one, but
the tachyon momenta are only constrained to verify (\ref{closed kinematic
shell}). This result can be extended to the four-tachyon of type 0 theory and
coincides with the one computed in \cite{KT}.

In conclusion we have computed projective invariant off-shell tachyon
amplitudes showing that, with a special choice of the local coordinate systems
around the punctures they can be put in the same form as their on-shell
counterparts. This has been done for the bosonic string and also for the
fermionic string, where little is known from the side of string field theory.
Suitable low-energy limits may allow to shed new light on effective actions
for tachyons and hence on their dynamics.

\bigskip

We acknowledge L. Cappiello, P.\ Di Vecchia and A. Liccardo for useful discussions.

\end{document}